\documentclass[aps,prd,10pt,twocolumn,superscriptaddress,showpacs,footinbib]{revtex4-1}
\pdfoutput=1
\usepackage{hyperref}
\usepackage{graphicx}
\usepackage{amsfonts,amsmath,amssymb,bm,bbm,setspace, booktabs, tabularx, multirow}
\usepackage{color}

\hypersetup{
    pdfnewwindow=true,      
    colorlinks=true,       
    linkcolor=blue,          
    citecolor=blue,        
    filecolor=blue,      
    urlcolor=blue        
}

\newcommand{\br}[0]{\mathbf{r}}
\newcommand{\los}[0]{\mathrm{LOS}}
\newcommand{\units}[1]{~\mathrm{#1}}
\newcommand{\comments}[1]{}

\graphicspath{{../img/}}

\begin{document}

\title{The Doppler effect on indirect detection of dark matter using dark matter only simulations}
\author{Devon Powell}
\email{dmpowel1@stanford.edu}
\affiliation{Kavli Institute for Particle Astrophysics and Cosmology (KIPAC),
	\\ Department of Physics, Stanford University, Stanford, CA 94305, USA}
\affiliation{SLAC National Accelerator Laboratory, Menlo Park, CA 94025, USA}
\author{Ranjan Laha}
\email{rlaha@stanford.edu}
\affiliation{Kavli Institute for Particle Astrophysics and Cosmology (KIPAC),
	\\ Department of Physics, Stanford University, Stanford, CA 94305, USA}
\affiliation{SLAC National Accelerator Laboratory, Menlo Park, CA 94025, USA}
\author{Kenny C. Y. Ng}
\email{chun-yu.ng@weizmann.ac.il}
\affiliation{Department of Particle Physics and Astrophysics, Weizmann Institute of Science, Rehovot 76100, Israel} 
\author{Tom Abel}
\affiliation{Kavli Institute for Particle Astrophysics and Cosmology (KIPAC),
	\\ Department of Physics, Stanford University, Stanford, CA 94305, USA}
\affiliation{SLAC National Accelerator Laboratory, Menlo Park, CA 94025, USA}
\date{\today}

\begin{abstract}
Indirect detection of dark matter is a major avenue for discovery.  However, baryonic backgrounds
are diverse enough to mimic many possible signatures of dark matter.  In this work, we study the newly proposed technique of dark matter velocity spectroscopy\,\cite{speckhard2016}.
The non-rotating dark matter halo and the Solar motion produce a distinct longitudinal dependence of
the signal which is opposite in direction to that produced by baryons.  Using collisionless dark
matter only simulations of Milky Way like halos, we show that this new signature is robust and holds great
promise.  We develop mock observations by high energy resolution X-ray spectrometer on a sounding rocket, the Micro-X experiment,  to our test case, the 3.5 keV line.  We show that by using six different pointings, Micro-X can exclude a constant line energy over various longitudes at $\geq$ 3$\sigma$.  The halo triaxiality is an important effect and it will typically reduce the significance of this signal.  We emphasize that this new {\it smoking gun in motion} signature of dark matter is general, and is applicable to any dark matter candidate which produces a sharp photon feature in annihilation or decay.
\end{abstract}


\maketitle

\section{Introduction}
\label{sec:Introduction}

The search for the particle properties of dark matter is one of the most important research avenues\,\cite{Jungman:1995df,Bertone:2004pz,Strigari:2013iaa}.  The ``weak" interactions experienced by the dark matter particle complicates these searches.  Despite decades of multi-pronged searches, we have not yet identified the dark matter particle\,\cite{Bertone:2016nfn}.  One of the most important ways to search for dark matter particles is indirect detection\,\cite{Klasen:2015uma,Gaskins:2016cha}.

Many anomalous signals have been interpreted as arising from dark matter interactions\,\cite{Elsaesser:2004ap,Loewenstein:2009cm,Prokhorov:2010us,Weniger:2012tx,Huang:2013pda,Gordon:2013vta,Daylan:2014rsa,Abazajian:2014hsa,Lee:2015fea,Bartels:2015aea,Bulbul:2014sua,Boyarsky:2014jta,Urban:2014yda}.  Astrophysical sources and detector artifacts have been known to mimic a dark matter signal\,\cite{Ackermann:2013uma,O'Leary:2015gfa,Ackermann:2015lka,Brandt:2015ula,O'Leary:2016osi,Gu:2015gqm,Phillips:2015wla,Shah:2016efh}. The separation of signal and background is difficult since one needs to model these in the same data set.  Taking lessons from all these misadventures, it is prudent to ask for new methods to cleanly separate the signal and background.  This is especially important since there are many viable dark matter candidates which can only be detected via astrophysical observations\,\cite{Bozek:2015bdo,Horiuchi:2015qri}.  

Distinct kinematic signatures arising from dark matter annihilation or decay are used to separate the dark matter signal from background.  These signatures include monochromatic photons arising from dark matter annihilation or decay.  Past experiences have shown that it is not reliable to only depend on this kinematic end point signature for the identification of a dark matter signal.  This raises the question: can we devise a new test of dark matter to confirm indirect detection signals?

Ref.\,\cite{speckhard2016} devised a new {\it smoking gun in motion} test of dark matter in indirect detection.  It utilized the superb energy resolution, $\sim \mathcal{O}$(0.1\%), of Hitomi (previously known as Astro-H)\,\cite{2014SPIE.9144E..25T} to find the new signature --- dark matter velocity spectroscopy.  Solar motion around the Galaxy produces a distinct longitudinal dependence in the dark matter signal, a signature of Doppler effect.  This new signature is model-independent and applicable to any dark matter signal containing a sharp feature.  It is unlikely that baryonic phenomena can produce such a distinct signature\,\cite{speckhard2016}.

Given the importance of dark matter particle searches, it is important to characterize any new model-independent signature in detail.  While Hitomi had a narrow field of view (FoV), it is important to confirm if dark matter velocity spectroscopy can also be performed by a high energy resolution instrument with a wide field of view.  These instruments can give strong and complementary limits on sterile neutrino dark matter decay\,\cite{McCammon:2002gb,Boyarsky:2006hr,Figueroa-Feliciano:2015gwa}.  In this work, we study the potential of dark matter velocity spectroscopy with a high energy resolution and wide FoV instrument, namely Micro-X.

We perform such a study in this work using collisionless dark matter only simulations from Mao etal.\,\cite{mao2015}.  Dark matter simulations take into account many complex non-linear processes which take part in galaxy formation.  This also helps us to take into account many halo properties which are not described by an analytical description of the halo. 

As an example of the dark matter signal, we consider the 3.5 keV line\,\cite{Bulbul:2014sua,Boyarsky:2014jta}.  The status of the 3.5 keV line is controversial\,\cite{Riemer-Sorensen:2014yda,Jeltema:2014qfa,Boyarsky:2014ska,Malyshev:2014xqa,Alvarez:2014gua,Anderson:2014tza,Boyarsky:2014paa,Bulbul:2014ala,Conlon:2014wna,Iakubovskyi:2014yxa,Carlson:2014lla,Jeltema:2014mla,Tamura:2014mta,Iakubovskyi:2015dna,Sekiya:2015jsa,Iakubovskyi:2015kwa,Riemer-Sorensen:2015kqa,Iakubovskyi:2015wma,Jeltema:2015mee,Ruchayskiy:2015onc,Bulbul:2016yop,Aharonian:2016gzq,Hofmann:2016urz,Arguelles:2016uwb,Conlon:2016lxl,Neronov:2016wdd,Perez:2016tcq,Gewering-Peine:2016yoj}.  The malfunctioning of the Hitomi satellite did not permit an observation to conclusively test this signal.  We use future Micro-X observations\,\cite{Figueroa-Feliciano:2015gwa} to demonstrate our technique.  It is expected that Micro-X will have an energy resolution (FWHM) of 3 eV at 3.5 keV\,\cite{Figueroa-Feliciano:2015gwa}, a high enough energy resolution to permit dark matter velocity spectroscopy\,\cite{speckhard2016}.  We emphasize that we are using this 3.5 keV signal as a proxy, and that the underlying physics of dark matter velocity spectroscopy is particle physics model-independent. 

There have been many works in which velocity spectroscopy was used to understand baryonic astrophysical emission\,\cite{Dame:2000sp,Diehl:2006cf,Kalberla:2008uu,Kretschmer:2013naa}.  Ref.\,\cite{speckhard2016} first applied this technique analytically to dark matter (see the footnote mentioning private communication by J. Bahcall in\,\cite{Turner:1986vr}).  In this work, we analyze for the {\it first} time dark matter velocity spectroscopy using collisionless dark matter only simulations 

Any telescope with $\mathcal{O}$(0.1\%) energy resolution can perform dark matter velocity spectroscopy.  An improvement in the energy resolution is the natural step in the evolution of telescope instrumentation.  This improvement will help in disentangling the dark matter signal from background, and improve our knowledge of the astronomical sources.  For certain wavelengths, it is already known how to build a detector with $\mathcal{O}$(0.1\%) energy resolution, such as INTEGRAL/ SPI\,\cite{2003AA}, XQC\,\cite{Figueroa-Feliciano:2015gwa}, and Hitomi\,\cite{Takahashi:2012jn,2014SPIE.9144E..25T}.  Near future instruments like Micro-X\,\cite{Figueroa-Feliciano:2015gwa}, ATHENA X-IFU\,\cite{Barret:2016ett}, and HERD\,\cite{Wang:2015ema,Huang:2015fca} will also have $\sim \mathcal{O}$(0.1\%) energy resolution.


\section{Theory}
\label{sec:theory}

In this section, we outline the theoretical insights leading to dark matter velocity spectroscopy\,\cite{speckhard2016}.  
The discussion is tailored for a wide field of view instrument like Micro-X.  In this work, we are concerned with sterile neutrino, 
$\nu_s$, decay to an active neutrino, $\nu_a$, and a photon, $\gamma$: $\nu_s \rightarrow \nu_a + \gamma$.  
This implies that the photon energy $E = m_s/2$.  We concentrate on the detection of photons in this work.

\subsection{Derivation}
\label{sec:derivation}

We now derive the main analytic expressions needed for velocity spectroscopy. For notational
convenience, we define the position vector $\br=\br(s, \Omega)$, where $s$ is the distance
along the line of sight, and $\Omega$ is the solid angle.  For instruments with an energy resolution $\delta E/E$ $\gg$ $\mathcal{O} (0.1\%)$, the differential flux of photons originating from dark matter decay in the Milky Way halo is given by\,\cite{Figueroa-Feliciano:2015gwa}:
\begin{eqnarray}
\dfrac{d^2 \mathcal{F}}{d\Omega \, dE} =  \dfrac{\Gamma}{4\pi \, m_s} \, \dfrac{dN}{dE} \, \int
_0 ^{s_{\rm max}}  \, ds \, \rho(\br)  \,.
\label{eq:double differential for the flux}
\end{eqnarray}
Here $\mathcal{F}$ denotes the flux in cm$^{-2}$ s$^{-1}$, $\Omega$ denotes the solid angle in sr,
$E$ denotes the energy of the photon in keV, and $\Gamma$ denotes the decay rate (in s$^{-1}$) of
the dark matter particle of mass $m_s$ (in keV).  The dark matter density (in keV cm$^{-3}$) profile
is denoted by $\rho(\br)$, and the photon spectrum (in keV$^{-1}$) is denoted by $dN(E)/dE$.  The line of sight distance, $s$, varies from 0 to $s_{\rm max}$, where the maximum value of the line of sight distance, $s_{\rm max}$, approximately corresponds to the virial radius of the Milky Way halo.

Observation by a telescope with $\sim$ $\mathcal{O}$(0.1\%) energy resolution modifies Eqn.\,\ref{eq:double differential for the flux} in two important and distinct ways.  First, the photon line is broadened due to the velocity dispersion of the dark matter particles in the Milky Way halo.  Second, the photon energy experiences Doppler shift.  

\begin{figure}[!h]
\centering
\includegraphics[width=0.99\columnwidth]{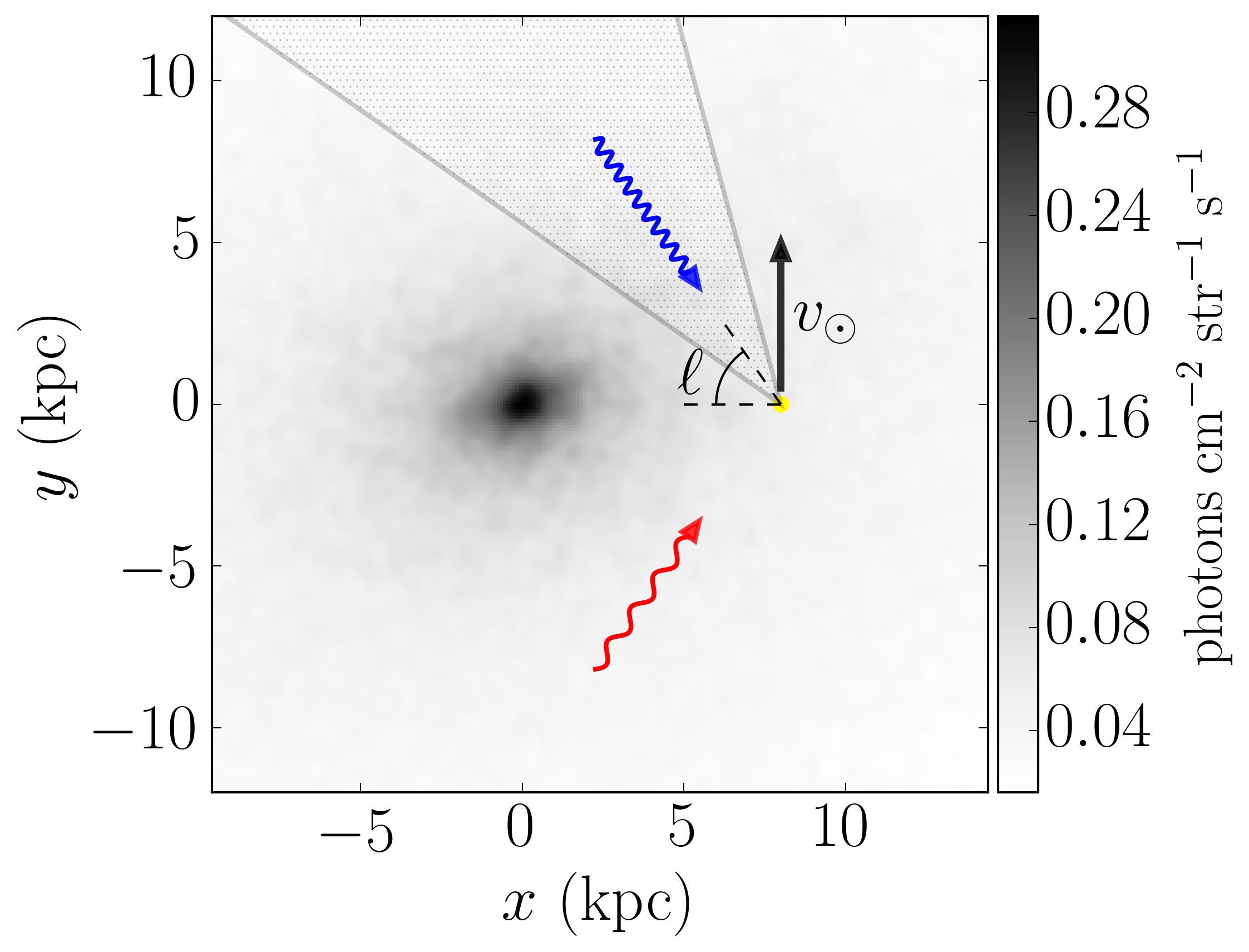}
\caption{The projected column density of Halo 374\,\cite{mao2015}, the most spherical halo in the suite of	simulations used here. The color scale is given in units of the surface brightness observed at the distance of the Sun, 8 kpc from the halo center. The diagram illustrates the basic principle of velocity spectroscopy: X-ray photons due to sterile neutrino dark matter decay are collected from within the field of view (shaded triangle) at the Sun's position. The net blue- or redshift of observed photons is determined by the line of sight relative to the Sun's velocity. In this simple model, the energy shift $\delta E/E=(v_\odot/c)\,\sin\ell\,\cos b$ (see Sec.\,\ref{sec:derivation}).}
\label{fig:halo374}
\end{figure}

We take into account the broadening of the line by convolving $dN(E)/dE$ by a Gaussian of width
$\sigma_E = (E/c) \sigma_{v_\los}$\,\cite{speckhard2016}.  Here $\sigma_{v_\los}(\br)$ 
is the line of sight velocity dispersion of dark matter.  The Gaussian form arises 
since we consider a Maxwellian dark matter velocity distribution.  The line shape will slightly change if we consider dark matter velocity distribution favored by recent hydrodynamical simulations\,\cite{Ling:2009eh,Kuhlen:2013tra,Butsky:2015pya,Bozorgnia:2016ogo,Sloane:2016kyi,Kelso:2016qqj}, but we ignore this small difference in this work.  

The broadened line spectrum can be written as 
\begin{eqnarray}
	\dfrac{d \tilde{N} (E, \br)}{dE} &=& \int dE' \, \frac{dN(E)}{dE}  G(E - E', \sigma_{E'} (\br)) \, ,
\label{eq:formula for modified dNdE}
\end{eqnarray}
where the convolution function is a Gaussian, as mentioned above.  The width of the Gaussian,
$\sigma_E(\br)$, is calculated following Ref.\,\cite{speckhard2016}.

Since the solar velocity $v_\odot \ll c$, we can use the nonrelativistic formula for Doppler shift: $\delta
E/E = - v_{\rm LOS}/c$.  Following Ref.\,\cite{speckhard2016}, we define $v_{\rm LOS} = (\langle
{\bf v}_\chi \rangle - {\bf v}_\odot) \cdot \hat{r}_{\rm LOS}$.  We assume $\langle {\bf v}_\chi
\rangle$ $\approx$ 0, and $v_\odot$ = 220 km s$^{-1}$.  We ignore smaller peculiar velocities, $\sim \mathcal{O}$(10 km s$^{-1}$), in this expression\,\cite{speckhard2016}.  In the coordinate system where the x-axis is
towards the Galactic Center, the direction of the Galactic rotation is in the y direction, and the
z-axis is normal to the Galactic plane, we have ${\bf v}_\odot$ = $v_\odot \, \hat{y}$.  In this
reference frame, ${\bf v}_\odot \cdot \hat{r}_{\rm LOS}$ = $v_\odot y / |\bf{r}_{\rm LOS}|$.  In terms of
the Galactic longitude, $\ell$, and Galactic latitude, $b$, we have $y = r_{\rm LOS} \, {\rm sin} \,
\ell \, {\rm cos} \, b$.  From this, we have $v_\los = v_\odot \sin \ell\cos b$.

Taking these two effects into account, we can rewrite Eqn.\,\ref{eq:double differential for the flux} as 
\begin{eqnarray}
\dfrac{d^2 \mathcal{F}}{d\Omega \, dE} &=&  \dfrac{\Gamma}{4\pi \, m_s} \, \int _0 ^{s_{\rm
max}}  \, ds \, \rho(\br) \nonumber\\
&\times& \dfrac{d \tilde{N}[E(1 - v_\los(\br)/c), \br]}{dE} \,.
\label{eq:double differential for the flux rewritten}
\end{eqnarray}
An important difference between the narrow field of view and wide field of view instruments is
encapsulated in the term $E(1 - v_\los(\br)/c)$.  For Hitomi, with a field of view of 9 arcmin$^2$, the maximum value of the line intensity approximately occurs at the center of the field of view.  This is not true for a wide field of view instrument like Micro-X (20$^\circ$ radius field of view).  The maximum value of the line intensity depends on the density profile as is evident from Eqn.\,\ref{eq:double differential for the flux rewritten}.  The line is also wider when compared to that observed in a narrow field of view instrument.    

The shift in the central value of the energy of the widened line is shown by the argument $(E/c)
v_\los(\br) $.  Both the width and observed energy of the line are determined by the position
$\br$ of the emission, indicated by $\br$ in the argument of $d\tilde{N}/dE$.

\subsection{Analytic model}
\label{sec:anmod}


Here we describe the analytic model to which we compare the results of our N-body analysis.
For our purposes, we assume that $dN/dE$ is a line. Thus, $d\tilde{N}/dE$ is a Gaussian of width $\sigma_E$
in accordance with Eqn. \eqref{eq:formula for modified dNdE}.

We begin with the total flux, which is found simply by integrating the J-factor over the field of
view. 
\begin{eqnarray}
\mathcal{F} =  \frac{\Gamma}{4\pi \, m_s} \, \int _\Omega \, \int _0 ^{s_{\rm max}}  \,  d\Omega\,ds
\, \rho(\br) \, .
\label{eq:lineflux}
\end{eqnarray}
In the limit of a small FoV, the J-factor can be assumed constant over the FoV. As we
consider a large FoV instrument here, we retain the full form of this integral in our calculations.

The observed Doppler shift is given in terms of the LOS velocity, weighted by the flux:
\begin{eqnarray}
	\langle v_\los\rangle &=& \frac{1}{\mathcal{F}}\frac{\Gamma}{4\pi \, m_s} 
	\int _\Omega \, \int _0 ^{s_{\rm max}}  \, d\Omega\,ds \, \rho(\br) \, v_\los(\br) \, .
\label{eq:lineshift}
\end{eqnarray}
Again, in the limit of a small FoV, $\langle v_\los\rangle = v_\odot \sin \ell\cos b$. We refer to this hereafter
as the ``simple sinusoid model''. 

Finally, the Doppler-broadened width of the line can be expressed simply when $d\tilde{N}/dE$ is a Gaussian:
\begin{eqnarray}
	\langle \sigma_{v_\los, {\rm total}}^2 \rangle &=& \frac{1}{\mathcal{F}}\frac{\Gamma}{4\pi \, m_s} 
	\int _\Omega \, \int _0 ^{s_{\rm max}}  \, d\Omega\,ds \, \rho(\br)\nonumber \\ 
	&\times&   [(v_\los(\br)-\langle v_\los\rangle)^2+\sigma_{v_\los}^2(\br)] \, .
\label{eq:lineshift}
\end{eqnarray}
Here the first term arises from the slightly different shifts in the FoV, whereas the second term is due to intrinsic width.  We compute these integrals analytically using a standard numerical quadrature routine.

\section{Methods}

\subsection{Simulations}
\label{sec:simulations}

Numerical simulations take into account many different processes which participate in dark matter halo formation.  Many signatures of these non linear processes are not taken into account in an analytical model of the dark matter halo.  It is thus important to validate any new signature of dark matter by using simulations of galaxy formation.

We evaluate the potential of dark matter velocity spectroscopy using collisionless dark-matter-only N-body
simulations.  We study a suite of Milky Way analogues run using the L-GADGET cosmology code (a
descendant of GADGET-2\,\cite{springel2005}). These are dark-matter-only zoom-in simulations run by
Ref.\,\cite{mao2015} to study subhalo abundance. Their high resolution and multiple realizations
makes them suitable for our purposes as well. Each halo has $\mathcal{O}(10^7)$ high-resolution
particles with a particle mass $m_p=4.0 \times 10^5 M_{\odot}$ and total  mass
$M_{\mathrm{vir}}\simeq 10^{12} M_{\odot}$ (masses in physical units). Refer to \cite{mao2015} for
the full description of the simulation parameters.

A $7\units{keV}$ sterile neutrino, produced by the Shi-Fuller mechanism\,\cite{Shi:1998km}, has a non-negligible free-streaming length and hence introduces a cutoff in
the matter power spectrum at a wave number $k_\mathrm{WDM} \simeq 10 \units{Mpc^{-1}\,h}$
\cite{ven2016}. This gives a corresponding mass scale cutoff of $M_\mathrm{WDM} \simeq 10^{10}
\units{M_\odot}$, or roughly $10^{-2}$ of the Milky Way halo mass. As we are interested in the main
halo itself, we do not concern ourself with the differences in smaller mass scales that may
arise due to our use of pure CDM simulations as a proxy for sterile neutrino dark matter. 
It has been shown that the flux maps for cold versus keV-mass dark matter
haloes differ by a few percent at most\,\cite{Lovell:2014lea}. While we focus on $m_s=7\units{keV}$ here,
one should view this study as a test model while noting that the velocity spectroscopy approach is
valid over a wide range of particle masses.  A different production mechanism for sterile neutrinos can change the free streaming length\,\cite{Merle:2014xpa}, but the general idea of velocity spectroscopy is valid for {\it any} production mechanism of dark matter.

When using simulations, the modeled DM decay signal is not isotropic about the Galactic Center.
Therefore, we must also explicitly set the observer's position and orientation when performing a
synthetic observation. For the majority of the work, we are agnostic to the
distribution of the dark matter prior to an observation. When setting up each synthetic observation, we simply generate a
new random orientation on the sphere using a uniform sampling algorithm (see e.g. Ref. \cite{arvo1992}) and
then apply that rotation to the particle data. The exceptions are the sky maps and the study of
the effects of halo triaxility (Section \ref{sec:results}), for which observer positions were
predetermined.

Among the 46 halo realizations, we focus on one halo, Halo 374, (Figure\,\ref{fig:halo374}) in detail. This is the most spherically-symmetric halo, with principal axis ratios $b/a=0.86$ and $c/a=0.73$.  We choose this halo to facilitate the comparison of our results with a spherically symmetric generalized NFW profile of the halo.  Dark-matter-only simulations produce triaxial halos, but recent hydrodynamical simulations have shown that the inclusion of baryons tend to sphericalize a halo\,\cite{Debattista:2007yz,Bryan:2012mw,Bernal:2014mmt,Bernal:2016guq}.

Although hydrodynamical models are investigated by many groups, it is not yet known if all the baryonic processes are self consistently taken into account in these simulations.  To take into account these uncertainties, we will also show our results for halos which are less spherical compared to Halo 374.  The statistical significance of the result depends on the triaxiality of the halo.

\subsection{Velocity spectroscopy using simulations}
\label{sec:simulations}

In order to generalize the analytical methods presented in Sec.\,\ref{sec:theory} to a simulation
containing N-body particles, we simply convert the integration to a sum over the N-body particles.
This is similar in spirit to the ``sightline'' method employed by \cite{Lovell:2014lea} and the
velocity distribution function sampling of \cite{Mao:2012hf}.  We construct the full spectral
intensity seen by the detector directly from the N-body particles, incorporating Doppler shift and
velocity dispersion in a natural way.   


The total flux can be found by integrating the differential flux in Eqn.\,\ref{eq:double differential for the flux} over the energy and solid angle.  Implementing this in an N-body simulation implies a summation over all of the particles, $p$, within the field of view, $\Omega$, and weighting by the inverse square of the scalar distance to the observer, $r^{-2}_p$:
\begin{eqnarray} 
\mathcal{F} = \frac{\Gamma}{4\pi \, m_s} \, \sum_{p \, \in \, \Omega} \, \frac{m_p}{r_p^{2}} \, .
\end{eqnarray}

The differential flux in energy can also be calculated in a similar way:
\begin{eqnarray}
\label{eq:discrete}
\frac{d\mathcal{F}}{dE} = \frac{\Gamma}{4 \pi \, m_s}\, \sum_{p \, \in \, \Omega} \,
\frac{m_p}{r_p^{2}} \, \frac{dN[E(1-v_p/c)]}{dE} \, ,
\end{eqnarray}
where $v_p$ is the velocity of particle $p$ projected along the line of sight to the observer.  By
considering the LOS velocity of each particle independently, we automatically capture the spectral
convolution introduced by the bulk velocity dispersion. 

\begin{figure}[h!]
\centering
\includegraphics[width=1.0\columnwidth]{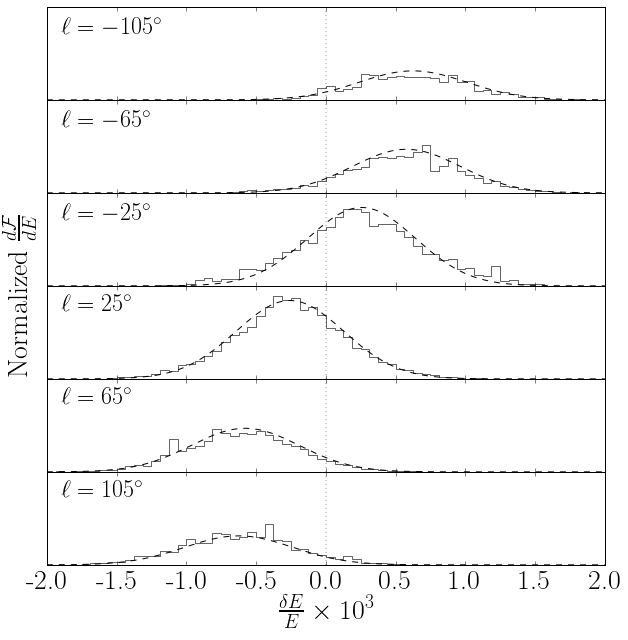}
\caption{We plot the decay line spectrum $d\mathcal{F}/dE$ as a function of fractional shift in the energy $\delta
E/E$ for six different pointings $\ell$, with $b=25^\circ$. This compares the empirical histograms
(solid) computed from the N-body simulation against the analytic Gaussian model (dashed) computed from an
NFW profile with an analytic velocity dispersion model. The
	Gaussian line profile agrees closely with the empirical spectrum. The data for all six pointings are
normalized relative to one another, and the x-axis is multiplied by 1000 for clarity. 
}
\label{fig:dfde}
\end{figure}

We focus here on the special case where $dN/dE$ is a line due to sterile neutrino decay.  The
parameters of the sterile neutrino are those favored by the 3.5 keV line\,\cite{Bulbul:2014sua}.  In
this case, computing the observed spectrum is then as simple as building a flux-weighted histogram
of the line-of-sight velocities for all particles in the sampling cone. In practice we find
$d\mathcal{F}/dE$ to be very nearly Gaussian. 
We compare the line width and shift computed analytically and directly from simulations in
Fig.\,\ref{fig:dfde} for six different fields of view at $b=25^\circ$.  The analytical computation is 
shown by dashed line, whereas the solid line shows the lines computed directly from the simulations.  
Good agreement is seen between these two different computations for this halo.


\subsection{Synthetic observation by Micro-X}
\label{sec:microx}

The instrument that we focus on in this work is Micro-X\,\cite{Figueroa-Feliciano:2015gwa}.  It is a
direct successor to the XQC sounding rocket
experiment\,\cite{McCammon:2002gb,Boyarsky:2006hr,Crowder:2012ts}.  We follow the parameters of the
instrument near 3.5 keV as mentioned in\,\cite{Figueroa-Feliciano:2015gwa}: 20$^\circ$ radius field
of view, effective area 1 cm$^2$ , 96\% detector efficiency, 300 s exposure time, and 3 eV
energy resolution (FWHM). For convenience we combine these instrumental parameters into the exposure factor
$O_{MX}=288\units{cm^2~s}$.

To minimize contamination by the Galactic plane, we focus on fields of view which are centered at $b = 25^\circ$ for various different Galactic longitudes, $\ell$.  This choice is a compromise between the decreasing signal strength away from the Galactic Center, and the more rapidly decreasing background away from the Galactic plane.  A typical sterile neutrino decay flux at $3.5$ keV from
the Milky Way halo at $b=25^\circ$ is $\mathcal{F}\sim 0.1$ photons\,cm$^{-2}$\,str$^{-1}$\,s$^{-1}$ 
for a signal count $N_s \sim 3-12$ photons, depending on $\ell$. 

We model the background $N_b$ using the cosmic X-ray background model of \cite{Ajello:2008xb}.  The
use of an alternative model\,\cite{2009A&A...493..501M} for the cosmic X-ray background increases the background rate by $\sim$ 20\%, but does not alter our
conclusions significantly.  

In order to understand the impact of point sources in our fields of view, we use the MAXI/ SSC catalog\,\cite{Tomida:2016ntt}.  The catalog gives the time averaged flux of X-ray sources in the energy range 0.7 - 7 keV separated into two energy bands: soft (0.7 - 1.85 keV) and hard (1.85 - 7 keV).  We assume that the time averaged X-ray flux $dN/dE \, = \, N_X \, E^\alpha$  (in units of keV$^{-1}$ cm$^{-2}$ s$^{-1}$), so that the flux measured by MAXI/ SSC in the soft and hard band is given by $\int _{0.7 \, {\rm keV}}^{1.85 \, {\rm keV}} E (dN/ dE) \, dE$ and $\int _{1.85 \, {\rm keV}}^{7 \, {\rm keV}} E (dN/ dE) \, dE$ respectively.  We can estimate the normalisation and the spectral index of the flux for each source using the measurements in these two bands.  The values of the normalisations and the spectral indices for the sources in our fields of view is given in Table\,\ref{tab:presentyields}.  Due to the large field of view, the cosmic X-ray background produces the larger contribution, and the contribution of the point sources is small ($\lesssim$ 20\%).  We are only interested in the flux in a small energy range around 3.5 keV, and the exact value of the spectral index does not play a significant role.  There are a few other sources in the MAXI/ SSC catalog which fall in our fields of view.  For these sources, Ref.\,\cite{Tomida:2016ntt} did not mention the flux in either the soft or the hard energy band, and thus we were unable to infer the normalization and spectral index using our method outlined above.  We expect that these sources will have a negligible contribution to the background counts.  

\begin{table}[b]

\caption{The normalisation and the spectral index of the various sources which fall inside the fields of view chosen in this paper.  The sources are taken from the MAXI/ SSC catalog\,\cite{Tomida:2016ntt}.}
\setlength{\extrarowheight}{4pt} 
\begin{ruledtabular}
\begin{spacing}{1.1}
\begin{tabular}{lcc}
Source & $N_X$ (in keV$^{-1}$ cm$^{-2}$ s$^{-1}$) & $\alpha$ \\ 
\hline
1ES 1959+650 & 0.04 & - 2.11 \\
Abell 2319 & 0.04 & -1.92 \\
3C 390.3 & 0.007 & -1.22 \\
Swift J1753.5-0127 & 0.2 & 1.97 \\
GX 9+9 & 0.59 & -1.18 \\
Abell 2256 & 0.016 & -1.49 \\
Her X-1 & 0.016 & -0.95 \\
Mrk 501 & 0.058 & -1.91 \\
Abell 2199 & 0.025 & -1.33 \\
Abell 2147 & 0.025 & -2.06 \\
SN 1006 & 0.044 & -2.26 \\
EX Hya & 0.034 & -1.62 \\
Centaurus Cluster & 0.073 & -1.87 \\
Abell 1060 & 0.019 & -1.55 \\
MAXI J0918-121 & 0.015 & -1.73 \\
Abell 754 & 0.021 & -1.48 \\
Abell 644 & 0.015 & -1.55 \\
\end{tabular}
\end{spacing}
\end{ruledtabular}
\label{tab:presentyields}
\end{table}

Since we have only assumed a time averaged spectrum, the background can be higher if a source is in a flaring state, but that can be tackled only with a real observation.  The photon energy range depends on the pointing of Micro-X.  We find that the range 3.5 keV $\pm$ 10.5 eV is wide enough to cover the signal for all the pointings considered in this work.
 This range contains all of the expected signal and captures
sufficient background counts for signal and background to be differentiated effectively. The
background spectrum is constant to within 1\% over this small range, so we fit a flat spectrum in
our analysis. A typical background count in this energy window is $N_b \sim 3$ photons per pointing.

Our synthetic observation itself consists of a Monte Carlo sampling of photons from the N-body particles
and the background model. For each particle $p \in \Omega$ in the sampling cone, we sample $n_p \sim
\mathrm{Poisson}(\lambda_p)$ photons, where the expected photon count per particle
$\lambda_p=\frac{\Gamma O_\mathrm{MX}m_p}{4\pi m_s r_p^2}$.
Each signal photon is given energy $E_i \sim \mathcal{N}[E(1-v_p/c),
\sigma_\mathrm{instr}]$ to model the energy resolution of the instrument. In
addition, we sample $n_b \sim \mathrm{Poisson}(\lambda_b)$ background photons, where the rate $\lambda_b$ is
determined by the background model \cite{Ajello:2008xb}, and endow each with energy
$E_i \sim \mathrm{Uniform}(3.5\units{keV}\pm10.5\units{eV})$. 

We then attempt to reconstruct the sterile neutrino decay line from the synthetic photons $\{E_i\}$.
A major challenge in this work is differentiating between small numbers of signal v/s background photons in
order to determine the line energy. To this end, we use an extended maximum likelihood analysis for
each pointing $\ell$.  This is an unbinned analysis described by \cite{barlow1990} which allows us to estimate
the signal parameters $E$ and $\sigma_E$ as well as the signal and background counts $N_s$
and $N_b$ simultaneously. This technique has been used by Ref.\,\cite{Figueroa-Feliciano:2015gwa} as well, as it gives
good fits to unbinned data for small number counts. 

\begin{figure}[h!]
\centering
\includegraphics[width=1.0\columnwidth]{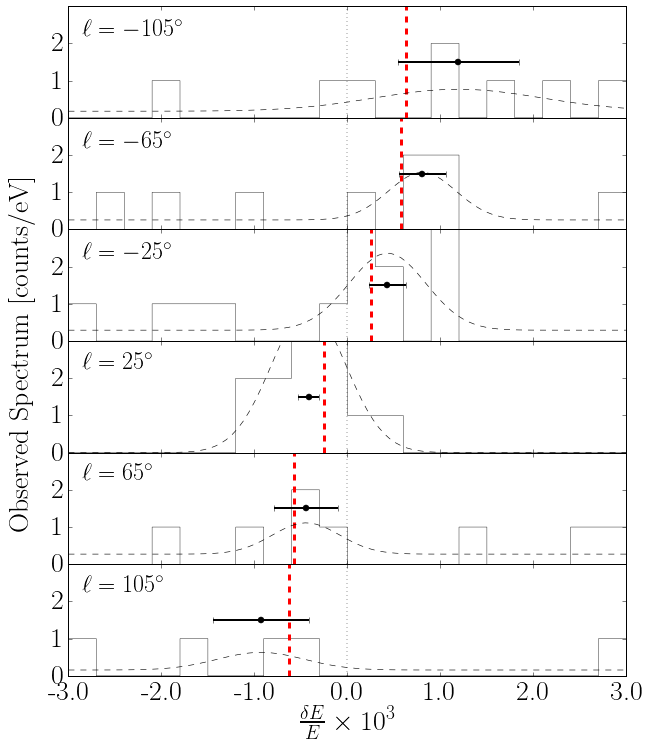}
\caption{ An example of a synthetic Micro-X observation. The histogram gives the observed photons,
	though we emphasize that this is only to guide the eye; the likelihood analysis itself is
	unbinned. The dashed black curves show the model fit, with the horizontal error bars giving the
	position of the line centroid and $1\sigma$ uncertainty. This uncertainty is computed via the
	extended likelihood analysis described by Ref. \cite{barlow1990}. The vertical red dashed lines give the line
	energy as predicted by the analytic model.
}
\label{fig:synobs}
\end{figure}

Our likelihood function is
\begin{multline}
	\mathcal{L}(E, \sigma_E, N_s, N_b; \{E_i\}) =\\
	\frac{e^{-(N_s+N_b)}}{N!} \prod_{i=1}^{N}
	\left(\frac{N_s e^{\frac{-(E_i-E)^2}{2(\sigma_E^2+\sigma_\mathrm{instr}^2)}}}
	{\sqrt{2\pi(\sigma_E^2+\sigma_\mathrm{instr}^2)}}+\frac{N_b}{\sigma_b}\right) \, .
\end{multline}
Fixed parameters are the total number
of observed photons $N$, the energy range over which the likelihood is fit $\sigma_b=21 \units{eV}$, 
and the Gaussian equivalent instrumental energy resolution $\sigma_\mathrm{instr}=1.3 \units{eV}$. The use of $\sigma_\mathrm{instr}$
in our fit models the uncertainty in the observed photon energies $E_i$ and serves to regularize the problem.
We give an example of one such synthetic observation and extended likelihood fit in Figure
\ref{fig:synobs}.


Our figure of merit for a detection of a Doppler-shifted line is the probability that the data
exclude zero shifting. In other words, we consider the energy shift of the line centroid
away from $\delta E/E=0$ in units of $\sigma_\mathrm{cent}$, the uncertainty in the line energy.

Ref.\,\cite{barlow1990} gives a prescription for obtaining the covariance matrix between the four fitted
parameters, which we use to compute $\sigma_\mathrm{cent}$. This naturally incorporates the Poisson errors due to
$N_s$ and $N_b$, so that $\sigma_\mathrm{cent} \sim \left(\frac{\sigma_E^2+\sigma_\mathrm{instr}^2
}{N_s}\right)^{1/2}$ as expected. 
We obtain a global significance by summing the significances (the
number of $\sigma_\mathrm{cent}$ by which $\delta E/E=0$ is excluded) for each
pointing in quadrature.

\section{Results}
\label{sec:results}

Our velocity spectroscopy analysis on N-body data reveals three main insights.

\begin{figure}[h!]
\centering
\includegraphics[width=1.0\columnwidth]{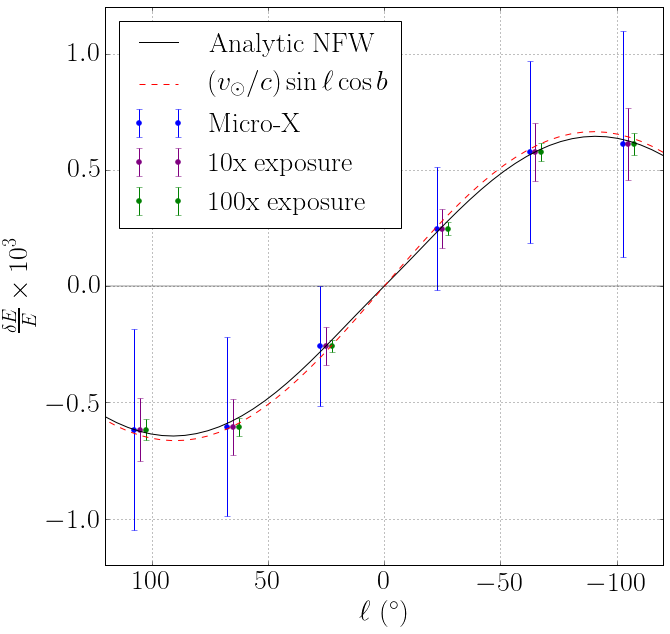}
\caption{Velocity spectroscopy on the Milky Way analogue Halo 374. This shows the position of the observed line centroid due to sterile
	neutrino decay as a function of Galactic longitude $\ell$, with latitude $b=25^\circ$.
	Here we compare results as computed from the N-body simulation (see Section \ref{sec:simulations}) for the
	instrumental parameters of Micro-X with our analytic model (Section
	\ref{sec:theory}) as well as a simple sinusoidal model, showing excellent agreement between the
	three. Note that the error bars represent $\sigma_{\mathrm{cent}}$, the $1\sigma$ uncertainty
	in the energy of the line centroid. This uncertainty is dominated by the Poisson statistics; we
	see that for hypothetical exposures with 10x and 100x the photon counts, the error bars shrink
	considerably. This also illustrates that the analytic NFW model (Section \ref{sec:anmod}), the small-FOV
	sinusoidal model, and the N-body data are indistinguishable until a 100x exposuire is
	considered. For the
	purposes of this plot and the sky maps (figures \ref{fig:skymaps} and \ref{fig:triax}), we estimate
	$\sigma_\mathrm{cent} = (\sigma_E^2+\sigma_\mathrm{instr}^2)^{1/2}C(N_b/N_s)N_s^{-1/2}$, where
	$C(R)=\sqrt{1+4R}$ is a factor given by the optimal Cramer-Rao bound.}
\label{fig:de_vs_l}
\end{figure}

\begin{figure}[h!]
\centering
\includegraphics[width=1.0\columnwidth]{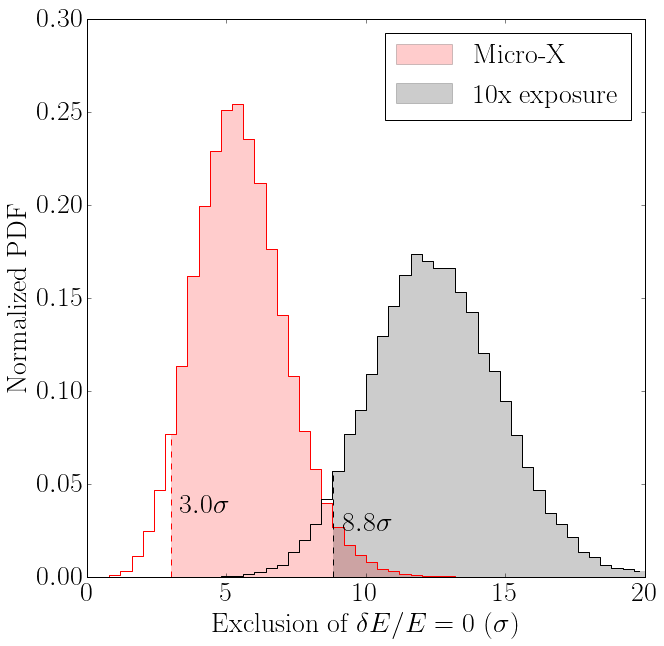}
\caption{ 
	PDFs of the detection significance for Doppler shifting of the dark matter decay line, given in
	units of $\sigma$. Vertical dashed lines and shaded regions give the 95\% confidence interval.
	We see that for the fiducial Micro-X parameters (red), 95\% of observations allow us to exclude
	$\delta E/E=0$ globally by $3\sigma$ or greater. If we model a hypothetical observing mission with
	10x the exposure time (gray), we find that the detection significance increases drastically, giving an
	$8.8\sigma$ detection in 95\% of observations. 
	These PDFs were sampled from $10^5$ synthetic Micro-X observations. Observations were randomly distributed
	between all haloes in the simulation suite, with the relative orientation of the Sun to the halo
	randomized as well. 
}
\label{fig:sigma_pdf}
\end{figure}

The first is that our analytic model matches the N-body calculation
extremely well  ($\chi^2_\mathrm{red}=1.8\times 10^{-3}$).  We summarize this result in Figure
\ref{fig:de_vs_l}. Furthermore, the simple sinusoid model given by $\delta E/E
= (v_\odot/c) \sin \ell \cos b$ (corresponding to the limit in which the field of view radius goes to
zero) also matches the analytic model for Micro-X to within a few percent over the entire range of
$\ell$ ($\chi^2_\mathrm{red}=5.7\times 10^{-3}$). This line-of-sight integral is thus
a sufficient approximation to a large field of view instrument for the given exposure. 
Note that such small $\chi^2_\mathrm{red}$ values are due to the large error bars from small $N_s$.
Since $\chi^2_\mathrm{red} \propto N_s$, these two models are 
only differentiable by observations $\sim 100$ times longer.  
This suggests that for further studies, this sinusoid model is more than sufficient and may be
used to quickly explore the parameter space of haloes (e.g. with a Markov chain Monte Carlo) with minimal loss of precision.

The second main result is that combining the observed line energies from the six pointings modeled
here ($\ell=\pm25^\circ,\pm65^\circ,\pm105^\circ$ with $b=25^\circ$) can exclude
$\delta E/E = 0$ globally at $\geq 3\sigma$.  A constant line shift over various longitudes imply a detector artifact, and hence the significance of excluding a baryonic origin of the line is even higher.  In $10^5$ synthetic observations as described in
Section \ref{sec:microx}, we find that 95\% allow a $3\sigma$ detection of Doppler shifting. Repeating this
experiment for a mission with 10x the exposure, we find that 95\% of observations give a $>8\sigma$
exclusion of $\delta E/E = 0$. See Figure \ref{fig:sigma_pdf} for the full results of this test.

\begin{figure}[h!]
\centering
\includegraphics[width=1.0\columnwidth]{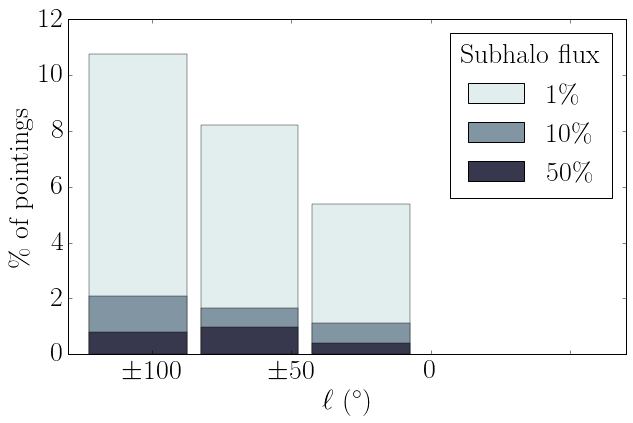}
\caption{ 
	The contribution of subhaloes to the total observed DM decay flux. 
	The bars give the fraction of all Micro-X pointings containing a subhalo that
	contributes more than some fraction of the total flux, i.e., we find that at $\ell=\pm105^\circ$,
	about 10\% of pointings contain a subhalo bright enough to contribute over 1\% of the total flux
	for that pointing. The number of pointings dominated by (10\% or 50\% of total flux) a single subhalo is low, at
	the $\sim$1\% level. This plot is symmetric in $\ell$, so only half of the pointings are plotted
	here.
}
\label{fig:subhalo_flux}
\end{figure}

We also find that the effect of subhaloes on our observations are small,
with only $\sim1\%$ of pointings seeing over 10\% of the flux contributed by a single subhalo
(Figure \ref{fig:subhalo_flux}). This supports intuition in that the perigee of subhaloes to the
galactic center in our
simulations is much larger than $r_\odot$: Only 1\% of subhaloes can be found within
$21\units{kpc}$ of the GC, so the smooth component of the main halo is expected to dominate
the signal.

\begin{figure}[h!]
\centering
	\includegraphics[width=0.99\columnwidth]{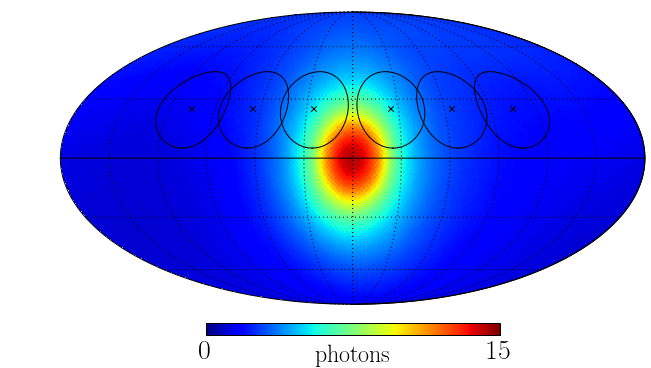}
	\includegraphics[width=0.99\columnwidth]{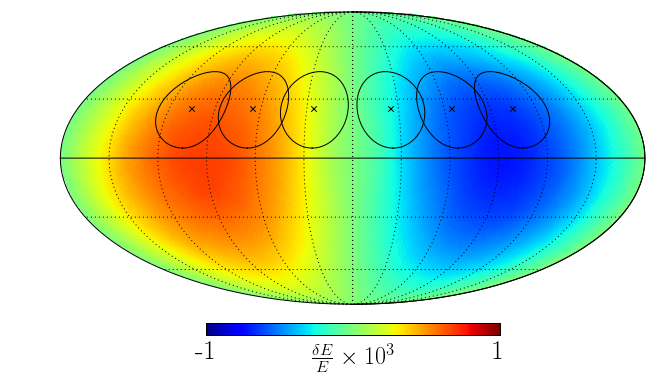}
	\includegraphics[width=0.99\columnwidth]{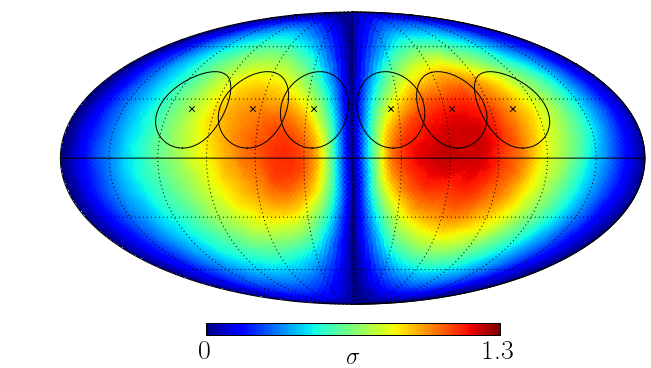}
\caption{Sky maps illustrating the principle of velocity spectroscopy for our Micro-X observation on
	Halo 374. Top: The flux map. Middle:
	The dipole pattern in the line energy induced by our motion relative to the Milky Way halo.
	Bottom: The significance of the detection of Doppler shifting of a dark matter decay line,
indicating the exclusion of$\delta E/E=0$ by a Micro-X
observation. The maps shown are actually the convolution of the sky brightness with the
$20^\circ$-radius field of view. Therefore, the observed flux, Doppler shift, and significance are
given by the color at the small X's.
Black circles indicate the
FoV of Micro-X on the sky for the six pointings considered here. }
\label{fig:skymaps}
\end{figure}

The pointings $\ell$ centered around $\pm75^\circ$ exclude $\delta E/E=0$ with 
the highest significance, at $\sim 1.5\sigma$.  This pointing optimizes between two
competing effects. The first is the higher J-factor, hence higher
photon flux, nearer the Galactic center, giving a smaller uncertainty in the energy of a line
detected at small $\ell$. The second is that the mean velocity along the line-of-sight relative to
the dark matter increases with $\ell$ (up to $\ell=90^\circ$), shifting the observed line further
from $\delta E/E=0$. We illustrate this effect using sky maps in Figure \ref{fig:skymaps}.
Future observations should focus on these longitudes
in order to achieve the best potential detection of this Doppler shift.

\begin{figure*}[!thpb]
\centering
	\includegraphics[width=0.49\textwidth]{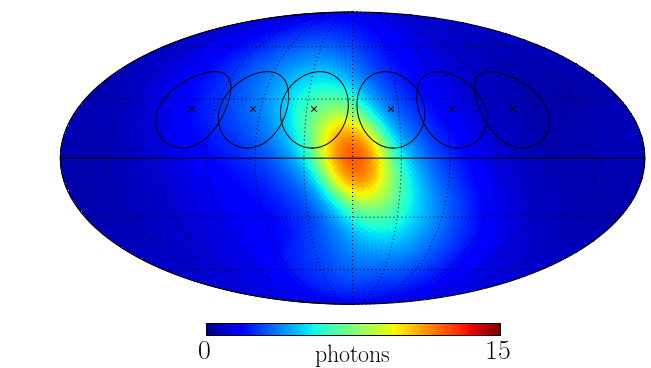}
	\includegraphics[width=0.49\textwidth]{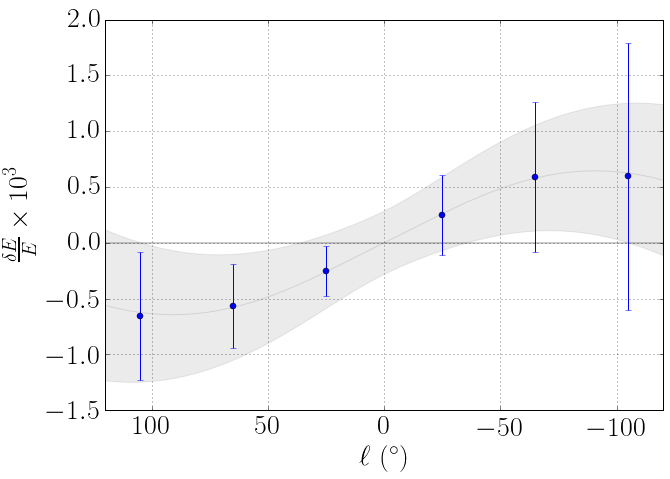}
\caption{An asymmetric Milky Way halo (Halo 800 with $b/a = 0.60$ and $c/a = 0.42$) can skew the significance of the
	observed signal to the east or west. Each field of view (black circles in the sky map where $b = 25^\circ$) corresponds to a data point in
the right hand plot. The gray shaded region indicates the $1\sigma$ uncertainties for the analytic model
derived from a spherical NFW profile fit to this halo.}
\label{fig:triax}
\end{figure*}

Finally, we find that halo triaxiality can bias the significance of a detection to
the east or west of the Galactic center. When observing above the Galactic plane (in this work,
$b=25^\circ$) any ellipticity of the Galactic halo on the sky can give a higher flux to one side of
$\ell=0^\circ$ than the other. While the mean prediciton for the line centroid matches the analytic model
for a spherical halo quite well, observing to one side of the GC allows one to exclude $\delta
E/E=0$ with higher significance due to better photon counting statistics. This is illustrated in
Figure \ref{fig:triax}. 

\section{Conclusions}
\label{sec:conclusions}

Numerous times in the past, anomalous astrophysical signals have been interpreted as the signature of dark matter.  So far, none of these extraordinary claims survived the scrutiny of extraordinary evidence.  All of these false alarms raise an important question:  can we design a new test to confirm the dark matter origin of an anomalous astrophysical signal?  

For the case of a photon line, the answer is yes.  Ref.\,\cite{speckhard2016} showed that telescopes with $\mathcal{O}$(0.1\%) energy resolution can utilize the Doppler effect of a sharp signal arising from dark matter annihilation and decay to perform a particle physics model-independent test: dark matter velocity spectroscopy.

We look into the issue in this work using synthetic observations generated from a suite
of N-body simulations \cite{mao2015}.  An N-body simulation takes into account many complex processes which occur during galaxy formation, and these effects are not completely captured in an analytical description.  It is essential to check the validity of any new signature with  numerical simulations.  We find that the phenomenon of dark matter velocity spectroscopy is robust while considering various dark matter only simulations of the Milky Way.  Previous work had applied the technique to a narrow field of view instrument like Hitomi, whereas in this work we concentrate on a wide field of view instrument like Micro-X.  We take the 3.5 keV line as our test case.  We show that if Micro-X confirms the presence of the 3.5 keV line in our Galaxy, then it can also use dark matter velocity spectroscopy to confirm or refute the dark matter origin of the line.

 For a Micro-X like instrument, the simple sinusoidal model for the Doppler shifting signature $\delta E/E
= (v_\odot/c) \sin \ell \cos b$ is sufficient (see Fig.\,\ref{fig:de_vs_l}).  This agreement arises due to the small number of photon counts.  We find that Micro-X (total
exposure $\sim10^3\units{cm^2~str~s}$ for all pointings) can reliably confirm the Doppler
shifting effect on a sterile neutrino decay line at $3\sigma$ or better (see Figs.\,\ref{fig:sigma_pdf} and \ref{fig:skymaps}). The best design for such an
observation consists of pointings centered around Galactic longitude $\ell\sim\pm75^\circ$ and
latitude $b=25^\circ$, as this gives the optimal balance between the magnitude of the Doppler shift
and the signal to noise ratio. 

Typical sterile neutrino dark matter production mechanisms predict a large cutoff in the mass function, $\sim$10$^8$ M$_\odot$.  Thus, subhalos contribute little to the flux spectrum, and has no appreciable effect to velocity spectroscopy for sterile neutrino dark matter (see Fig.\,\ref{fig:subhalo_flux}).  Dark matter halos are predicted to be triaxial.  We find that triaxiality has a large effect on velocity spectroscopy: the Doppler shift becomes asymmetric, and it results in a decrease of the signal significance (see Fig.\,\ref{fig:triax}).

Since we look at regions away from the Galactic Center, the signal rate is not maximum at these places.  The observation strategy for any search should first focus on searching for a line at regions with higher signal to noise.  The technique of dark matter velocity spectroscopy should be then applied at this specific line energy to confirm or refute the dark matter interpretation of this photon line.

We are optimistic about future experiments. While Micro-X is a good model for a first observation of dark matter velocity
spectroscopy in the Milky Way, future high energy resolution telescopes are sure to provide the
modest increase in photon counts needed for a high-significance detection. Increasing the 
Micro-X exposure by a factor of 10 gives at least an $8\sigma$ detection of the Doppler
effect on photons arising from sterile neutrino decay.

The theoretical and
observational concepts laid out in this work are general.  Any sharp photon feature arising from dark matter annihilation or decay\,\cite{Boddy:2015efa,Ibarra:2013eda} will get Doppler shifted in the above mentioned way.  In addition to other indirect searches of dark matter (see for e.g.,\,\cite{Yuksel:2007dr,Dasgupta:2012bd,Laha:2012fg,Ng:2013xha,Murase:2015gea,Ng:2015gfa,Chowdhury:2016bxs}), we expect that this new technique will be adopted widely.  We encourage theorists, experimentalists and observers to optimize their tools to take advantage of this new and exciting {\it smoking gun in motion} signature of dark matter.

	

\section*{Acknowledgments} 

The authors thank John F. Beacom, Eric Charles, Joseph Conlon, Chris Davis, Matt Kistler, Dmytro Iakubovskyi, Mark Lovell, and Greg Madejski for helpful discussions.  We especially thank Yao-Yuan Mao for help with the simulations.  D.P. is
thankful for support from the Stanford Fletcher Jones Foundation Fellowship.  R.L. is supported by DOE Contract DE-AC02-76SF00515 in SLAC.



\newcommand{\mnras}[0]{M.N.R.A.S.}
\bibliography{references}	

\end{document}